\documentclass[%
 reprint,
superscriptaddress,
 amsmath,amssymb,
 aps,
]{revtex4-2}

\usepackage{graphicx}
\usepackage{dcolumn}
\usepackage{bm}
\usepackage[colorlinks=true, citecolor=blue, linkcolor=blue, urlcolor=blue]{hyperref} 
\usepackage{float}
\usepackage{placeins}

\begin{document}

\preprint{APS/123-QED}

\title{The effect of space charge on photon-enhanced thermionic emission \\ in the presence of the bidirectional discharge}

\author{Xinqiao Lin}
\affiliation{Department of Physics, Xiamen University, Xiamen 361005, People’s Republic of China}
\author{Ousi Pan}
\affiliation{Department of Physics, Xiamen University, Xiamen 361005, People’s Republic of China}
\author{Zhimin Yang}
\affiliation{School of Physics and Electronic Information, Yan’an University, Yan’an 716000, People’s Republic of China}
\author{Yanchao Zhang}
\affiliation{School of Science, Guangxi University of Science and Technology, Liuzhou 545006, People’s Republic of China}
\author{Jincan Chen}
\affiliation{Department of Physics, Xiamen University, Xiamen 361005, People’s Republic of China}
\author{Shanhe Su}
 \email{sushanhe@xmu.edu.cn}
\affiliation{Department of Physics, Xiamen University, Xiamen 361005, People’s Republic of China}


\date{October 24, 2024}

\begin{abstract}
The bidirectional space charge effects in photon-enhanced thermionic emission (PETE) devices are investigated systematically. First, we precisely determine the carrier concentrations and cathode temperatures by taking into account the electron recycling effect, energy balance constraints, and space charge effects arising from the concurrent discharge of the cathode and anode. Next, we analyze the impact of critical parameters, including anode properties and operating conditions, on the space charge barrier distribution and the overall performance of the device. The results demonstrate that the impact of reverse discharge on the net current becomes more pronounced when the PETE device operates at high anode temperatures, low anode work functions, and with a moderate solar concentration ratio and gap width. This discovery not only deepens our understanding of the bidirectional space charge effect, but also provides valuable guidance for the future optimization of PETE device performance.
\end{abstract}

\maketitle


\section{\label{sec:Introduction}Introduction}

Solar energy is widely acknowledged as one of the most promising avenues toward carbon neutrality, given its status as an inexhaustible, environmentally-friendly, and ubiquitous energy source \cite{Zhang_2022_AssessingEnergyTransition}. Over the past decades, the development of this renewable energy source has generated significant interest in high-efficiency solar cell technologies \cite{Guo_2024_QuantumInterferenceRecombination, Li_2024_FlexibleSiliconSolar, Odebowale_2024_DesignOptimizationNearfield}. Thermionic emission, unlike photovoltaics which rely on the photovoltaic effect, is a process that can directly convert heat energy into electricity and has seen significant advancements in recent years \cite{Jensen_2021_SubmicrometerGapThermionicPower, Chen_2022_PhysicsBasedModelNonuniform, Franey_2024_EnhancementThermionicPower}. Its current emission follows the Richardson-Dushman law \cite{Dushman_1923_ElectronEmissionMetals}, theoretically enabling conversion efficiencies over 30$\%$ in thermionic energy converters \cite{Campbell_2021_ProgressHighPower}. In 2010, Schwede et al. \cite{Schwede_2010_PhotonenhancedThermionicEmission} introduced a novel photon-enhanced thermionic emission (PETE) mechanism, which combines both the photoelectric and thermionic effects to enable photoelectric energy conversion under high-temperature conditions. This has garnered significant attention among the theoretical models of various emerging photoelectric converter technologies \cite{Schwede_2013_PhotonenhancedThermionicEmission, Rahman_2021_SemiconductorThermionicsNext}. 

This PETE mechanism has garnered significant attention. The theoretical conversion efficiency of PETE devices can surpass 40$\%$, and potentially reach up to 50$\%$ when coupled with a secondary heat engine in series \cite{Schwede_2010_PhotonenhancedThermionicEmission}. However, such PETE devices do not always operate at high efficiency in the photon-enhanced mode and may degrade to the pure thermionic emission mode \cite{Rahman_2021_SemiconductorThermionicsNext}, highlighting the importance of continued research and optimization for these devices.

The space charge effect poses a major challenge in the development of high-performance thermionic devices \cite{Hatsopoulos_1973_ThermionicEnergyConversion, Hatsopoulos_1979_ThermionicEnergyConversion}. Electrons emitted from the cathode/anode take time to reach the opposite electrode, accumulating as a negative charged cloud in the gap between the electrodes. This negative charged space charge cloud creates an additional potential barrier, hindering electron movement and reducing the net current \cite{Zhang_2021_SpaceChargeLimited, Zhang_2017_100YearsPhysics}. In the early 20th century, the pioneering works of Child \cite{Child_1911_DischargeHotCao} and Langmuir \cite{Langmuir_1913_EffectSpaceCharge} laid the foundational understanding of space charge effects. Their renowned Child–Langmuir (CL) law describes the maximum steady-state current that can be constrained by Poisson's equation in a planar electrode gap. Fry \cite{Fry_1921_ThermionicCurrentParallel} introduced an analogy comparing electron behavior within a planar gap to that of a steady-state, collision-free gas, which was later refined by Langmuir \cite{Langmuir_1923_EffectSpaceCharge}. Hatsopoulos and Gyftopoulos \cite{Hatsopoulos_1973_ThermionicEnergyConversion, Hatsopoulos_1979_ThermionicEnergyConversion} built upon the principles of thermionic conversion and presented the space charge effect in a more standardized and comprehensive format. Khoshaman et al. \cite{Khoshaman_2015_NanostructuredThermionicsConversion} introduced a novel limit-approaching method to calculate the critical points, effectively overcoming the challenge of low precision commonly associated with traditional calculation methods.

Our research group  \cite{Su_2014_SpaceChargeEffects, Wang_2019_OptimalDesignInterelectrode} as well as other groups  \cite{Segev_2015_NegativeSpaceCharge, Qiu_2022_PhotothermoelectricModelingPhotonenhanced, Rahman_2021_SemiconductorThermionicsNext, Wang_2024_EffectSpaceCharge} have used the space charge theory to investigate PETE devices. These works have indicated that reducing the electrode gap is a theoretically viable approach to overcoming the space charge effects in PETE devices \cite{Su_2014_SpaceChargeEffects, Segev_2015_NegativeSpaceCharge}. However, this method of reducing the electrode gap may be constrained by the dual challenges of high heat loss due to near-field thermal radiation effects, as well as the need for high-precision metrology at the micro- and nano-scale manufacturing levels \cite{Wang_2019_OptimalDesignInterelectrode, Rahman_2021_SemiconductorThermionicsNext}. An alternative effective method to suppress the negative space charge effects is through the technique of cation neutralization \cite{Ito_2012_OpticallyPumpedCesium, Wang_2024_EffectSpaceCharge}. Additionally, the selection and configuration of appropriate electrode materials (such as those with negative electron affinity) can also help mitigate the space charge effect to a certain degree \cite{Smith_2007_ConsiderationsHighperformanceThermionic, Smith_2013_IncreasingEfficiencyThermionic}. 

In fact, the electron currents from both the cathode and anode can contribute to the formation of the space charge barrier. Previous researches on PETE devices have often assumed that the contribution of the anode current to the space charge effect was negligible. However, this assumption holds true only under strict conditions, such as a low-temperature anode and a moderate anode work function. Typically, if these conditions are not met, the reverse current from the anode not only directly reduces the net current magnitude, but also introduces an additional space charge barrier that hinders the movement of electrons.
In the field of thermionic energy conversion,  Kniazzeh et al. \cite{Kniazzeh_1959_PotentialDistributionTwo} were pioneering in incorporating the effects of the reverse current potential barrier into the Poisson equation, and integrating it using the MIT-IBM 704 computer with dimensionless parameters. Dugan \cite{Dugan_1960_ContributionAnodeEmission} designed an iterative algorithm to determine the relationship between the space charge-limited current and operating voltage for specific parameter configurations. Hatsopoulos and Gyftopoulos \cite{Hatsopoulos_1979_ThermionicEnergyConversion} incorporated and refined the earlier approaches in a more standardized format in their monograph. Lee et al. \cite{Lee_2012_OptimalEmittercollectorGap} conducted further studies on the interplay between the space charge effect and the near-field radiant heat transfer. Khoshaman et al. \cite{Khoshaman_2016_SelfconsistentApproachAnalysis} proposed a self-consistent algorithm based on the particle tracking method and found that the electron number density around the two electrodes exhibits two maximum values.

The existing research above provides a key insight: under certain conditions, the potential barrier created by reverse discharge can significantly impact the performance of thermionic devices. Given that PETE devices may encounter similar operational scenarios, it becomes equally crucial to investigate the bidirectional space charge effect in these systems. This is particularly important for the isothermal configuration \cite{Segev_2013_HighPerformanceIsothermal}, where the temperatures of the anode and cathode are equalized. Additionally, it is critical for hybrid energy conversion systems \cite{Qiu_2022_PhotothermoelectricModelingPhotonenhanced}, which require integrating waste heat utilization elements in series with the high-temperature anode. Finally, while a configuration with a low anode work function can improve efficiency \cite{Liang_2015_ElectronThermionicEmission}, it also amplifies the effects of reverse discharge.

In this paper, we propose a solution to explore the bidirectional space charge effect in PETE devices. We discuss the scenarios where the electron emission phenomenon of the anode is particularly prominent, and the reverse current potential barrier cannot be ignored. The goal of this paper is to unravel the influence of the bidirectional space charge effect through rigorous numerical simulations, integrating it with other crucial factors such as electron recycling  \cite{Segev_2012_EfficiencyPhotonEnhanced, Segev_2015_NegativeSpaceCharge} and energy balance \cite{Su_2013_ParametricOptimumDesign}. The rest contents are organized as follows: In Sec.~\ref{sec:model}, the model of the PETE device in the presence of the bidirectional discharge is built. In Sec.~\ref{sec:Result}, the effect of bidirectional discharge barrier and the performance characteristics of the PETE device are revealed by numerical calculation. Finally, we conclude with Sec.~\ref{sec:Conclusions}, which summarizes the key insights gained from the study.

\section{\label{sec:model}Theoretical model}

\subsection{The schematic of photon-enhanced thermionic emission}

\begin{figure*}[htbp]
\centering 
\includegraphics[width=0.8\linewidth]{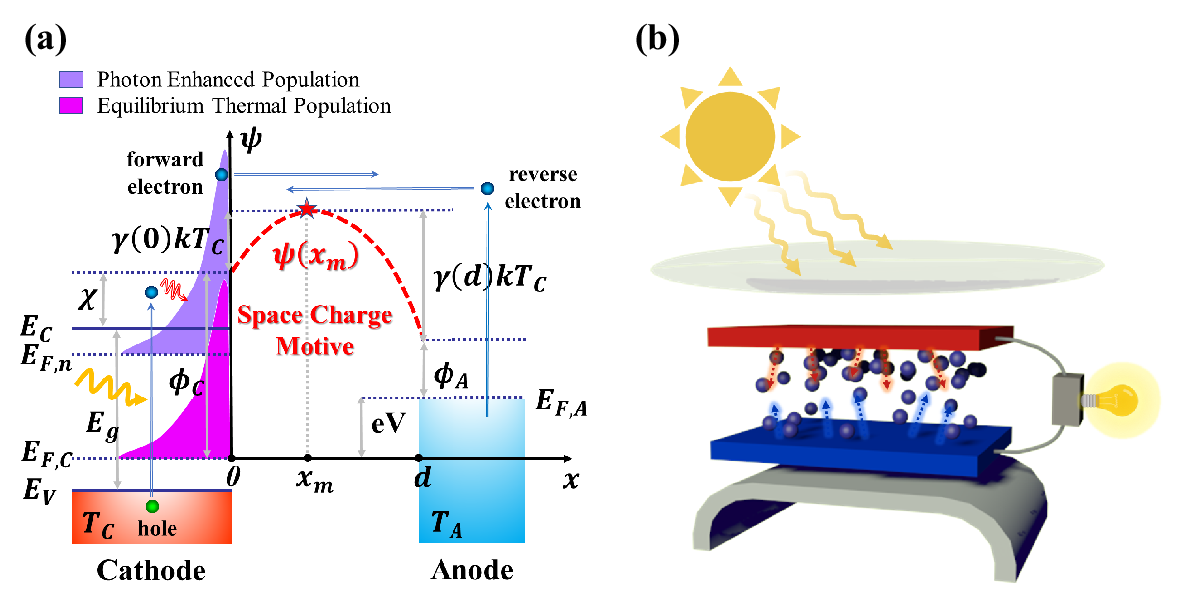} 
\caption{\label{fig:schematic}(a) The schematic diagram of photon-enhanced thermionic emission. (b) The implementation of a parallel-plate PETE device that converts solar energy into electrical energy through the photovoltaic effect and thermionic emission.}
\end{figure*}

Fig.~\ref{fig:schematic}(a) depicts the schematic diagram of the PETE device. It consists of a p-type semiconductor material serving as the cathode, and an anode metal plate with a vacuum gap in between. The blue straight arrow emanating from the hole (the green circle) represents the excitation of electrons from the valence band to the conduction band upon the absorption of photons (the yellow curved arrow). Following a rapid thermalization process, the excited electrons will disperse themselves uniformly across the entirety of the cathode material, attaining an equilibrium distribution in line with the temperature $T_C$ of the cathode. Electrons with energies exceeding the electron affinity $\chi$ can be emitted from the cathode surface, thereby generating a thermionic current. As a result, the emitted electrons can leverage the energy of the absorbed photons to surmount the material's bandgap, as well as the thermal energy necessary to overcome the material's electron affinity. As shown in Fig.~\ref{fig:schematic}(b), the cathode and anode form a parallel-plate structure. Thus, the surface areas available for both photon emission/absorption and electron emission/absorption are equal. Furthermore, the two plates are held at uniform temperatures, denoted as $T_C$ and $T_A$, respectively. Note that the cathode is assumed to be thermally integrated with an ideal solar-absorbing coating on its surface. This enables the cathode to absorb sub-bandgap photons as heat. Additionally, the cathode is presumed to have a uniform emissivity of unity across the entire spectrum.

\subsection{The charge neutrality without illumination }

A p-type semiconductor, such as boron-doped silicon, is used as the cathode material. The equilibrium Fermi level \(E_{F,C}\) of the semiconductor crystal under dark condition is approximated by solving the charge neutrality condition
\begin{equation}
{n_{eq}} + N_A^ -  = {p_{eq}}.
\label{eq:neutrality}
\end{equation}

The equilibrium concentrations of electron and hole carriers are determined as \({n_{eq}} = {N_C}{e^{ - \left( {{E_C} - {E_{F,C}}} \right)/k{T_C}}}\)  and \({p_{eq}} = {N_V}{e^{ - \left( {{E_{F,C}} - {E_V}} \right)/k{T_C}}}\), where \({E_C}\) and \({E_V}\)  are the energies of the conduction and valence band edges, respectively, and \({k}\) is the Boltzmann constant. The effective densities of states in the conduction band and valence band are represented by \({N_C}\) and \({N_V}\). These are calculated using the following relationships: \({N_C} = 2{\left( {2\pi m_n^*k{T_C}/{h^2}} \right)^{3/2}}\) and \({N_V} = 2{\left( {2\pi m_p^*k{T_C}/{h^2}} \right)^{3/2}}\), where \(m_n^* = {m_e}\) is the effective mass of the electron carriers, \(m_p^* = 0.57{m_e}\) is the effective mass of the hole carriers, \({m_e}\) is the mass of a free electron, and \(h\) is Planck's constant. Under the Boltzmann approximation, the concentration of ionized boron acceptors is expressed as \(N_A^ -  = {N_A}\left[ {\frac{1}{{1 + 4{e^{\left( {{E_A} - {E_{F,C}}} \right)/\left( {k{T_C}} \right)}}}}} \right]\), where \({E_A} = 0.044{\rm{eV}}\) is the ionisation energy of the boron acceptor in silicon, and the total concentration of the boron acceptor impurities \({N_A} = {10^{19}}{\rm{c}}{{\rm{m}}^{ - 3}}\). 

\subsection{The saturation current densities of the cathode and anode}

For the PETE device, the expression for the total saturation current density ${J_C}$ emitted out of the cathode is given by
\begin{equation}
{J_{SC}} = A{T_C}^2{e^{ - \frac{{{\phi _C} - \left( {{E_{F,n}} - {E_{F,C}}} \right)}}{{k{T_C}}}}},
\label{eq:J_SC}
\end{equation}
where the Richardson–Dushman constant \(A = 4\pi em_n^*{k^2}/{h^3}\), ${E_{F,n}}$ is quasi-Fermi level of electron in the cathode, and the work function ${\phi _C} = \chi  + {E_g} - {E_{F,C}}$. The term ${\phi _C} - \left( {{E_{F,n}} - {E_{F,C}}} \right)$ indicates that the effective work function is diminished by the discrepancy between the quasi-Fermi level ${E_{F,n}}$ under photoexcitation and the Fermi level ${E_{F,C}}$ in the absence of photoexcitation.

In non-degenerate semiconductors, the quasi-Fermi level ${E_{F,n}} = {E_{F,C}} + k{T_C}{\rm{ln}}\left( {\frac{n}{{{n_{eq}}}}} \right)$, where $n$ denotes the total electron concentration in the conduction band under photoexcitation conditions. Therefore, it is straightforward to rewrite Eq.~\eqref{eq:neutrality}  as
\begin{equation}
{J_{SC}} = en\sqrt {\frac{{k{T_C}}}{{2\pi {m_n}^*}}} {e^{ - \frac{\chi }{{k{T_C}}}}}.
\label{eq:J_SC_2}
\end{equation}

For the anode electrode, the saturation current density $J_{SA}$ is calculated by the traditional Richardson equation 
\begin{equation}
{J_{SA}} = A{T_A}^2\exp \left( { - \frac{{{\phi _A}}}{{k{T_A}}}} \right),
\label{eq:J_SA}
\end{equation}
where ${\phi _A}$ is the work function of the anode.

\subsection{The Child—Langmuir space charge theory in the presence of the bidirectional discharge}

When charge carriers move between the parallel-plane electrodes in a vacuum, they can form a space-charge region \cite{Langmuir_1923_EffectSpaceCharge}. The presence of space-charge effects can significantly limit the achievable current density of thermionic emission. To incorporate space-charge theory into the PETE, it is necessary to determine the electrostatic potential distribution in the interelectrode space, which requires the self-consistent solution of Poisson's equation \cite{Hatsopoulos_1979_ThermionicEnergyConversion}. By assuming that the electron flow is one-dimensional and collisionless, a motive diagram $\psi \left( x \right)$, as shown in Fig.~\ref{fig:schematic}(a), is used for the convenient analysis, where $x_m$ is the position in the interelectrode space. Because the motive $\psi \left( x \right)$ is the electrostatic potential times the negative value of the elementary charge $e$, the force on an electron is equal to the negative gradient of $\psi \left( x \right)$\cite{Hatsopoulos_1979_ThermionicEnergyConversion}. 
The Child—Langmuir space-charge theory, as applied in Ref.\cite{Hatsopoulos_1979_ThermionicEnergyConversion}, can be readily extended to analyze the motive distribution of vacuum diodes, where significant back emission is present. The motive diagram $\psi \left( x \right)$, which is proportional to the electrostatic potential, is determined by the solution to Poisson's equation in the region between the electrodes, i.e., 
\begin{equation}
\frac{{{d^2}\psi }}{{d{x^2}}} =  - \frac{{{e^2}N\left( x \right)}}{{{\varepsilon _0}}},
\label{eq:Poisson}
\end{equation}
where \({\varepsilon _0} = {\rm{8}}.{\rm{85}} \times {\rm{ 1}}{0^{ - {\rm{14}}}}{\rm{Fc}}{{\rm{m}}^{{\rm{ - 1}}}}\) is the permittivity of vacuum. The electron number density \(N\left( x \right)\) at position $x$ is obtained by integrating the electron distribution function \(f\left( {x,v} \right)\) over all the velocity components ${v_x}$, ${v_y}$, and ${v_z}$ in the Cartesian coordinate system, i.e.,
\begin{equation}
N(x) = \int_{ - \infty }^\infty  {d{v_y}\int_{ - \infty }^\infty  {d{v_z}\int_{ - \infty }^\infty  {d{v_x}f\left( {x,v} \right)} } }. 
\label{eq:N(x)}
\end{equation}

Note that \(f\left( {x,v} \right)\) depends on the position \(x\) and the magnitude of the velocity \(v = \sqrt {v_x^2 + v_y^2 + v_z^2} \) as well. At the location of the maximum motive \({x_m}\), the minimum value of the electron velocity \({v}\) is zero. The electron distribution function \(f\left( {x,v} \right)\) that describes the statistical behavior of the electrons in the interelectrode space can be modeled as the sum of two half-Maxwellian distribution functions. One of these two half-Maxwellian distributions corresponds to electrons that originate from the cathode and reaches the maximum motive \(\psi \left( {{x_m}} \right)\) with a minimum velocity ${v_x}$ equal to zero. All these electrons will pass the point of the maximum motive and be accelerated by the negative space charge towards the anode. The other half-Maxwellian distribution corresponds to electrons that originate from the anode and reach the maximum motive \(\psi \left( {{x_m}} \right)\) with a maximum velocity ${v_x}$ in the negative direction along ${x}$-axis equal to zero. All these electrons originating from the anode travel through the point of the maximum potential energy and are then accelerated by the negative space charge toward the cathode. It follows that the distribution function \(f\left( {{x_m},v} \right)\) at the maximum motive is given by 
\begin{equation}
f\left( {{x_m},v} \right) = {f_C}\left( {{x_m},v} \right) + {f_A}\left( {{x_m},v} \right), 
\label{eq:f(x_m,v)}
\end{equation}
where ${f_C}\left( {{x_m},v} \right) = \frac{{2N\left( {{x_m}} \right)}}{{1 + \alpha }}{\left( {\frac{{{m_e}}}{{2\pi k{T_C}}}} \right)^{\frac{3}{2}}}{e^{ - \frac{{{m_e}{v^2}}}{{2k{T_C}}}}}\tau \left( {{\nu _x}} \right)$, ${f_A}\left( {{x_m},v} \right) = \frac{{2\alpha N\left( {{x_m}} \right)}}{{1 + \alpha }}{\left( {\frac{{{m_e}}}{{2\pi k{T_A}}}} \right)^{\frac{3}{2}}}{e^{ - \frac{{{m_e}{v^2}}}{{2k{T_A}}}}}\tau \left( { - {\nu _x}} \right)$, and \(\tau \left( x \right)\) represents a Heaviside step function. $\alpha$ is the ratio of the electron number density at $x_m$ caused by electrons originating from the anode to that caused by electrons originating from the cathode, which is expressed as
\begin{equation}
\alpha  = \frac{{\int_{ - \infty }^0 d {v_x}\int_{ - \infty }^\infty  d {v_y}\int_{ - \infty }^\infty  d {v_z}f\left( {{x_m},v} \right)}}{{\int_0^\infty  d {v_x}\int_{ - \infty }^\infty  d {v_y}\int_{ - \infty }^\infty  d {v_z}f\left( {{x_m},v} \right)}}.
\label{eq:alpha_integral}
\end{equation}

For the case where the parameter $\alpha$ equals to zero, Eq.~\eqref{eq:f(x_m,v)} reduces to the regime of space-charge effect without back emission, which has been previously discussed in the literature on the PETE \cite{Su_2014_SpaceChargeEffects}.

We next examine the range of velocity values for the electrons emitted from the cathode. In order to make a better distinction, we defined the region between the surface of the cathode (\(x = 0\)) and the point $x_m$ of the maximum motive as the cathode space, and the region between the position of the maximum motive $x_m$ and the anode front surface at $x_d$ as the anode space. Each of these emitted electrons must overcome the maximum motive in order to reach the anode. In the cathode space, as shown in FIG.~\ref{fig:schematic}(a), the electrons are decelerated by the negative space charge. This is because the gradient of the motive is positive (\(d\psi /dx > 0\)) in this region. Electrons along $x$ direction that do not have enough initial energy to reach the point of maximum potential are returned to the cathode. As a result, the value of  $v_x$ can be either positive or negative. Electrons will have energies ranging from 0 to \(\psi \left( {{x_m}} \right) - \psi \left( x \right)\) as they return toward the cathode, and energies from 0 to infinity as they accelerate toward the anode. It follows that $v_x$ has a range of \( - {v_0} \le {v_x} < \infty \) \cite{Hatsopoulos_1979_ThermionicEnergyConversion}, where \({v_0} = {\left\{ {2\left[ {\psi ({x_m}) - \psi (x)} \right]/{m_e}} \right\}^{1/2}}\). In the anode space, as shown in FIG.~\ref{fig:schematic}(a), the electrons are accelerated by the negative space charge. This is because the gradient of the motive is negative (\(d\psi /dx < 0\)) in this region. The lowest energy electrons at position $x$ will have energy \(\psi \left( {{x_m}} \right) - \psi \left( x \right)\) toward the anode as they are accelerated toward the anode due to the electric field. The range of electron velocities $v_x$ at positions $x$ greater than $x_m$ is then given by \({v_0} \le {v_x} < \infty \) \cite{Hatsopoulos_1979_ThermionicEnergyConversion}. 

For the electrons originating from the anode, the range of velocity values at each position $x$ can be readily determined by using the same line of reasoning. Thus, we find that the actual range of electron velocities for the electrons originating from the anode is determined by
\[ - \infty  < {v_x} \le  - {v_0}\quad {\rm{ for\ }}x < {x_m}\]
  and
\[ - \infty  < {v_x} \le {v_0}\quad {\rm{ for\ }}x \ge {x_m}.\]

TABLE~\ref{table:v_x}  presents the range of velocities observed for the electrons emitting from different electrodes situated in various spatial regions. The table highlights the variability in electron velocities depending on the specific electrode and position in space.
\begin{table}[h]
\begin{ruledtabular}
\begin{tabular}{lcc}
Emission Source& \begin{tabular}{c}$v_x$ in the cathode\\ $(x < x_m)$\end{tabular} & \begin{tabular}{c}$v_x$ in the anode space\\ $(x \ge x_m)$\end{tabular} \\
\hline
Cathode& \(- {v_0} \le {v_x} < \infty \) & \({v_0} \le {v_x} < \infty \) \\
Anode& \( - \infty  < {v_x} \le  - {v_0}\) & \( - \infty  < {v_x} \le  {v_0}\) \\
\end{tabular}
\end{ruledtabular}
\caption{The range of electron velocities observed from the electrons emitting from different electrodes situated in various spatial regions.}
\label{table:v_x}
\end{table}

By considering Eq.~\eqref{eq:f(x_m,v)} as a boundary condition and combining it with the information provided in TABLE~\ref{table:v_x}, the general solution of the electron distribution function \(f\left( {x,v} \right)\) at any position $x$ can be solved by using the Vlasov equation \cite{Hatsopoulos_1979_ThermionicEnergyConversion}. The electron distribution function
\begin{equation}
f\left( {x,v} \right) = {f_C}\left( {x,v} \right) + {f_A}\left( {x,v} \right),
\label{eq:f(x,v)}
\end{equation}
where the distribution function \(f_C\left( {x,v} \right)\) corresponding to the electrons that originate from the cathode reads 
\begin{equation}
\begin{split}
{f_C}\left( {x,v} \right) &= \frac{{2N\left( {{x_m}} \right)}}{{1 + \alpha }}{\left( {\frac{{{m_e}}}{{2\pi k{T_C}}}} \right)^{\frac{3}{2}}}{e^{\gamma  - \frac{{{m_e}{v^2}}}{{2k{T_C}}}}} \\
&\times \left\{ {\begin{array}{*{20}{c}}
{\tau \left( {v + {v_0}} \right),{\rm{ }}x < {x_m}}\\
{\tau \left( {v - {v_0}} \right),{\rm{ }}x \ge {x_m}}
\end{array}} \right.,
\end{split}
\label{eq:f_C(x,v)}
\end{equation}
and the distribution function \(f_A\left( {x,v} \right)\) associated with the electrons that are emitted from the anode is expressed as 
\begin{equation}
\begin{split}
{f_A}\left( {x,v} \right) &= \frac{{2\alpha N\left( {{x_m}} \right)}}{{1 + \alpha }}{\left( {\frac{{{m_e}}}{{2\pi k{T_A}}}} \right)^{\frac{3}{2}}}{e^{\delta \gamma  - \frac{{{m_e}{v^2}}}{{2k{T_A}}}}} \\
&\times \left\{ {\begin{array}{*{20}{c}}
{\left[ { - \tau \left( {v + {v_0}} \right)} \right],{\rm{ }}x < {x_m}}\\
{\left[ { - \tau \left( {v - {v_0}} \right)} \right],{\rm{ }}x \ge {x_m}}
\end{array}} \right..
\end{split}
\label{eq:f_A(x,v)}
\end{equation}

In Eqs.~\eqref{eq:f_C(x,v)} and Eqs.~\eqref{eq:f_A(x,v)}, we define the dimensionless parameters as follows: \(\gamma (x) = [\psi ({x_m}) - \psi (x)]/(k{T_C})\) and $\delta  = {T_C}/{T_A}$.
Substituting Eqs.~\eqref{eq:f(x,v)}-~\eqref{eq:f_A(x,v)} into Eq.~\eqref{eq:N(x)} yields the electron number density \(N(x)\) as
\begin{equation}
\begin{split}
N\left( x \right) &= \frac{{N\left( {{x_m}} \right)}}{{1 + \alpha }} \\
&\left\{ {{e^\gamma }\left[ {1 \pm {\rm{erf}}\left( {\sqrt \gamma  } \right)} \right] + \alpha {e^{\delta \gamma }}\left[ {1 \mp {\rm{erf}}\left( {\sqrt {\delta \gamma } } \right)} \right]} \right\},
\end{split}
\label{eq:N(x)_2}
\end{equation}
where \({\rm{erf}}\left( u \right) = 2/\sqrt \pi  \int_0^u {{e^{ - {t^2}}}} dt\) is an error function. The upper signs in $ \pm $ and $ \mp $ are used for \(x < {x_m}\) and the lower signs are used for \(x \ge {x_m}\). When \(\alpha  = 0\), Eq.~\eqref{eq:N(x)_2} reduces to the corresponding equation derived for \(N\left( x \right)\) in Ref.\cite{Su_2014_SpaceChargeEffects}, where the space charge effect considers only the electrons emitted from the cathode.
By using Eq.~\eqref{eq:N(x)_2} and introducing the dimensionless distance variable \(\xi  = (x - {x_m})/{x_0}\) from the maximum motive, where \(x_0^2 = {\varepsilon _0}k{T_C}\left( {1 + \alpha } \right)/\left[ {2{e^2}N\left( {{x_m}} \right)} \right]\), the Poisson equation in Eq.~\eqref{eq:Poisson} can be expressed as \cite{Hatsopoulos_1979_ThermionicEnergyConversion}
\begin{equation}
2\frac{{{d^2}\gamma }}{{d{\xi ^2}}} = {e^\gamma }\left[ {1 \pm {\rm{erf}}\left( {\sqrt \gamma  } \right)} \right] + \alpha {e^{\delta \gamma }}\left[ {1 \mp {\rm{erf}}\left( {\sqrt {\delta \gamma } } \right)} \right],
\label{eq:Possion 2}
\end{equation}
where the upper signs in $ \pm $ and $ \mp $ are used for \(\xi  < 0\) and the lower signs are used for \(\xi  \ge 0\). As the point of \(\xi  = 0\) corresponds the position of the maximum motive, the boundary conditions \(\gamma {|_{\xi  = 0}} = 0\) and \(\frac{{d\gamma }}{{d\xi }}{|_{\xi  = 0}} = 0\) must be satisfied. Eq.~\eqref{eq:Possion 2} can be solved to get the implicit relation between \(\xi \) and \(\gamma \). This relationship can be converted to the motive distribution function $\psi \left( x \right)$ in the interelectrode space. 

\subsection{The net current density}

The net current density $J$ is the difference between the current density $J_C$ from the cathode and the current density $J_C$ from the anode, where the current densities are determined by the electrons going past the maximum motive \(\psi \left( {{x_m}} \right)\). Thus, the current density $J_C$ is given by the relation 
\begin{equation}
{J_C} = e\int_{ - \infty }^\infty  {d{v_y}\int_{ - \infty }^\infty  {d{v_z}\int_0^\infty  {d{v_x}{v_x}{f_C}\left( {{x_m},v} \right)} } }.
\label{eq:J_C intergal}
\end{equation}

To obtain the current density, $J_C$, one needs to solve Eq.~\eqref{eq:Possion 2} to get the motive diagram, and then the distribution function \({f_C}\left( {x,v} \right)\) as a function of position can be obtained from Eq.~\eqref{eq:f_C(x,v)}. Substituting the distribution function \({f_C}\left( {x_m,v} \right)\) at the maximum motive position $x_m$ into Eq.~\eqref{eq:J_C intergal}, $J_C$ can be integrated and expressed as:
\begin{equation}
{J_C} = e\frac{{N\left( {{x_m}} \right)}}{{1 + \alpha }}\sqrt {\frac{{2k{T_C}}}{{\pi {m_e}}}}.
\label{eq:J_C integrated}
\end{equation}

The saturation current \({J_{SC}}\), as described in Eq.~\eqref{eq:J_SC_2}, which is the initial emitted current from the cathode, can also be calculated by
\begin{equation}
{J_{SC}} = e\int_{ - \infty }^\infty  {d{v_y}\int_{ - \infty }^\infty  {d{v_z}\int_0^\infty  {d{v_x}{v_x}{f_C}\left( {0,v} \right)} } }.
\label{eq:J_SC intergal}
\end{equation}

By substituting Eq.~\eqref{eq:f_C(x,v)} into Eq.~\eqref{eq:J_SC intergal} and combining with Eq.~\eqref{eq:J_C integrated}, the relation between the current density \({J_C}\) from the cathode and the saturation current density \({J_{SC}}\) can be obtained as
\begin{equation}
{J_C} = {J_{SC}}\exp \left[ { - \gamma \left( 0 \right)} \right],
\label{eq:J_SC to J_C}
\end{equation}
where \(\gamma \left( 0 \right)\) represents the dimensionless difference between the maximum motive and the motive at the cathode surface.

According to the same calculation steps, the current density $J_A$ from the anode can be simplifies as
\begin{equation}
{J_A} = e\frac{{\alpha N\left( {{x_m}} \right)}}{{1 + \alpha }}\sqrt {\frac{{2k{T_A}}}{{\pi {m_e}}}},
\label{eq:J_A integrated}
\end{equation}
which is related to the saturation current density $J_{SA}$ in Eq.~\eqref{eq:J_SA} as
\begin{equation}
{J_A} = {J_{SA}}\exp \left[ { - \delta \gamma \left( d \right)} \right],
\label{eq:J_SA to J_A}
\end{equation}
where \(\gamma \left( d \right)\) represents the dimensionless difference between the maximum motive and the motive at the anode surface. By combing Eqs.~\eqref{eq:J_SC}, ~\eqref{eq:J_SA}, ~\eqref{eq:J_SC to J_C} and ~\eqref{eq:J_SA to J_A}, the net current density $J$ of the PETE device can be expressed as
\begin{equation}
\begin{split}
J &= {J_C} - {J_A} \\
&= AT_C^2{e^{ - \frac{{{\phi _C} - \left( {{E_{F,n}} - {E_{F,C}}} \right) + \gamma \left( 0 \right)k{T_C}}}{{k{T_C}}}}} - AT_A^2{e^{ - \frac{{{\phi _A} + \gamma \left( d \right)k{T_C}}}{{k{T_A}}}}}.
\end{split}
\label{eq:J=J_C-J_A}
\end{equation}

In addition, it is worth mentioning that, by combining Eqs.~\eqref{eq:J_C integrated} and \eqref{eq:J_A integrated}, the ratio $\alpha$ is demonstrated to be equal to $\frac{{{J_A}\sqrt {{T_C}} }}{{{J_C}\sqrt {{T_A}} }}$. Therefore, by selecting specific values for $J_C$, $J_A$, $T_C$, and $T_A$ and using a numerical method for ordinary differential equations, we can solve the Poisson equation [Eq.~\eqref{eq:Possion 2}] to obtain the distribution of $\gamma \left( x \right)$.

\subsection{The electron continuity equation in the conduction band}

From Eq.~\eqref{eq:J=J_C-J_A}, it is observed that the net current density relies on the quasi-Fermi level ${E_{F,n}}$ of electrons in the cathode. Before calculating ${E_{F,n}}$, the electron concentration $n$ must be determined by solving the following electron continuity equation in the conduction band
\begin{equation}
{\Gamma _{Sun}} - {\Gamma _R} = \frac{{{J_C} - {J_A}}}{eL}.
\label{eq:electron continuity}
\end{equation}

The generation rate of electrons ${\Gamma _{Sun}} = \frac{{\Phi _{Sun}}\left( {E > {E_g}} \right)}{L}$ \cite{Schwede_2010_PhotonenhancedThermionicEmission}, where ${\Phi _{Sun}}\left( {E > {E_g}} \right)$ is the photon flux density above the bandgap energy in the concentrated AM1.5 Direct (+circumsolar) solar spectrum, and $L$ is the film thickness. The rate of photon-enhanced radiative recombination is given by \({\Gamma _R} = \frac{1}{L}\left( {\frac{{np}}{{{n_{eq}}{p_{eq}}}}\; - 1} \right)\frac{{2\pi }}{{{h^3}{c^2}}}\mathop \smallint \limits_{{E_g}}^\infty  \frac{{{{(h\nu )}^2}d\left( {h\nu } \right)}}{{{e^{h\nu /\left( {k{T_C}} \right)}} - 1}}\)\cite{Schwede_2010_PhotonenhancedThermionicEmission}, where $c$ is the speed of light, $p$ is the concentration of hole in the valence band with photoexcitation, and $h\nu$ is the photon energy. For this analysis, we only consider radiative recombination as the recombination mechanism. Similar to Ref.\cite{Segev_2015_NegativeSpaceCharge}, Eq.~\eqref{eq:J_SC_2} assumes that both the reverse emission electrons and all electrons that are reflected by the energy barrier in the inter-electrode space contribute to the electron population, which demonstrates the electron recycling effect. The step-by-step derivation of $n$ from Eq.~\eqref{eq:electron continuity} can be found in the supporting materials from Ref.\cite{Segev_2015_NegativeSpaceCharge}.

\subsection{The energy balance at the cathode}

Furthermore, we analyze the energy balance constraints for the cathode, from which the cathode temperature can be determined. The radiative heat transfer between the cathode and the anode is neglected by assuming that the anode is perfectly reflective. By incorporating an infrared (IR) coupling element in the cathode to absorb sub-bandgap photons as heat, the energy balance equation of the cathode is
\begin{equation}
\begin{split}
{P_{sun}} &= {P_{IR}} + {P_0} + {P_{rad}} \\ 
&+ {J_C}\left[ {\psi \left( {{x_m}} \right) + 2k{T_C}} \right] - {J_A}\left[ {\psi \left( {{x_m}} \right) + 2k{T_A}} \right],
\end{split}
\label{eq:energy balance}
\end{equation}
where \({P_{sun}}\) is the input energy density of the concentrated AM1.5 Direct (+circumsolar) spectrum, ${P_0} = \frac{{2\pi }}{{{h^3}{c^2}}}\int_{Eg}^\infty  {\frac{{{{(hv)}^3}d(hv)}}{{{e^{hv/\left( {k{T_C}} \right)}} - 1}}} $ represents the loss of thermal energy due to the equilibrium radiative recombination of the cathode, ${P_{rad}} = {P_0}\left[ {np/\left( {{n_{eq}}{p_{eq}}} \right) - 1} \right]$ is the thermal energy loss due to the non-equilibrium radiative recombination, and ${P_{IR}} = \frac{{2\pi }}{{{h^3}{c^2}}}\int_0^{Eg} {\frac{{{{(hv)}^3}d(hv)}}{{{e^{hv/\left( {k{T_C}} \right)}} - 1}}} $ is the energy loss due to the radiation emitted from the IR coupling element. Note that the IR coupling element absorbs all sub-bandgap radiation and is fully transparent to supra-bandgap photons. The sum of ${P_{IR}}$ and ${P_0}$ represents the full-spectrum blackbody emission of the cathode. The last term ${J_C}\left[ {\psi \left( {{x_m}} \right) + 2k{T_C}} \right] - {J_A}\left[ {\psi \left( {{x_m}} \right) + 2k{T_A}} \right]$ in Eq.~\eqref{eq:energy balance} indicates the thermal energy that is carried away by the electrons emitted from the cathode and anode. Each electron emitted from the cathode to the anode carries away an energy $\psi \left( {{x_m}} \right) + 2k{T_C}$. Conversely, each electron flowing from the anode to the cathode gives up an energy $\psi \left( {{x_m}} \right) + 2k{T_A}$ as it arrives at the cathode \cite{Hatsopoulos_1973_ThermionicEnergyConversion, ODwyer_2007_SolidstateRefrigerationPowergeneration}.

\subsection{The power output and efficiency }

Finally, the voltage in the space charge limited regime can be written as 
\begin{equation}
V = \left\{ {\left[ {{\phi _C} + \gamma \left( 0 \right)k{T_C}} \right] - \left[ {{\phi _A} + \gamma \left( d \right)k{T_C}} \right]} \right\}/e.
\label{eq:V}
\end{equation}
The power output and efficiency of the PETE device can be expressed as
\begin{equation}
P = JV,
\label{eq:P}
\end{equation}
and
\begin{equation}
\eta = JV/{P_{sun}}.
\label{eq:eta}
\end{equation}

\section{\label{sec:Result}Results and discussion}

In the following discussion, we take AM1.5 Direct (+circumsolar) spectrum as the incident spectrum. If not otherwise specified, we set the solar concentrating ratio $C=500$, interelectrode gap width $d=5\mathrm{\mu m}$, energy band gap $E_g=1.4\mathrm{eV}$, electron affinity $\chi =0.4\mathrm{eV}$, and anode work function $\phi_A=0.9\mathrm{eV}$. To emphasize the effect of anode discharge, we set the anode temperature $T_A$ at a relatively large value, while the cathode temperature $T_C$ is determined by the energy balance equation [Eq.~\eqref{eq:energy balance}].

\begin{figure*}[htbp]
\centering
\includegraphics[width=0.8\linewidth]{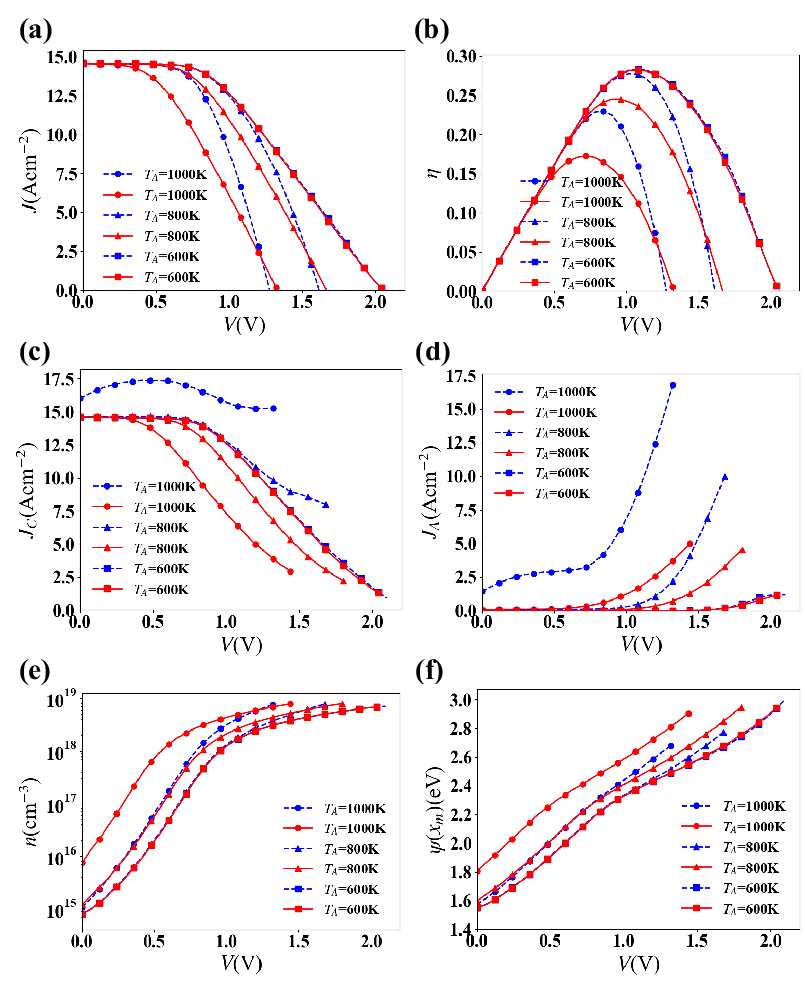} 
\caption{\label{fig:output}(a) The net current density $J$, (b) efficiency $\eta$, (c) current density $J_C$ from the cathode, (d) current density $J_A$ from the anode, (e) the electron concentration $n$ in the conduction band of the cathode, and (f) maximum motive $\psi \left( {{x_m}} \right)$ varying with the voltage $V$. The red lines correspond to the bidirectional discharge (BD) model, where the space charge effect is considered in the presence of the bidirectional discharge. The blue lines correspond to the forward discharge (FD) model, where the space charge effect is analyzed without considering the reverse current from the anode. The curves with circles, triangles, and squares are obtained for the anode temperature $T_A=1000\mathrm{K}$, $800\mathrm{K}$, and $600\mathrm{K}$, respectively.}
\end{figure*}

\begin{figure*}[htbp]
\centering
\includegraphics[width=0.8\linewidth]{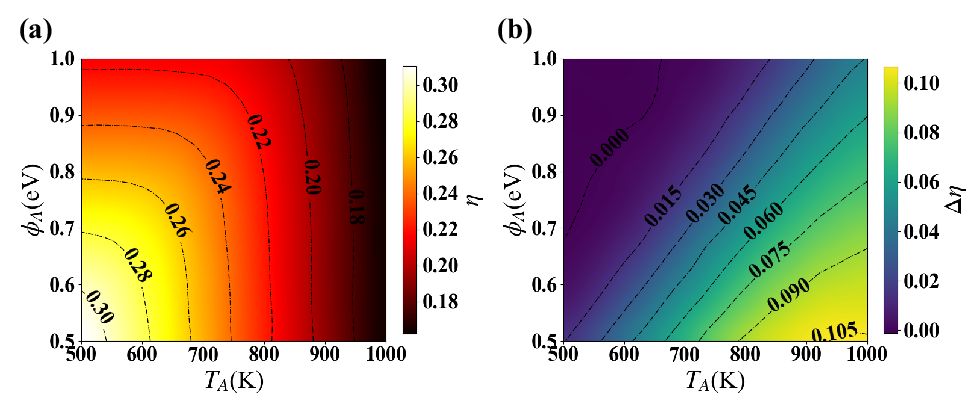} 
\caption{\label{fig:anode property}The contour plots of (a) the efficiency $\eta$ of the BD model and (b) the deviation $\Delta \eta$ of the efffieicny of the BD model with that of the FD model varying with the anode work function $\phi_A$ and the anode temperature $T_A$, where the voltage $V$ has been optimized.}
\end{figure*}

\begin{figure*}[htbp]
\centering
\includegraphics[width=0.8\linewidth]{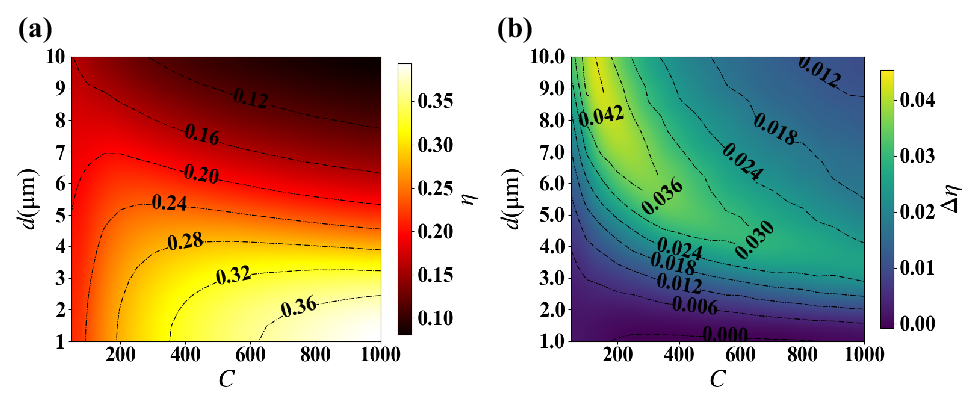} 
\caption{\label{fig:operating conditions}The contour plots of (a) the efficiency $\eta$ of the BD model and (b) the deviation $\Delta \eta$ of the efffieicny of the BD model with that of the FD model varying with the gap width $d$ and the solar concentration $C$, where the voltage $V$ has been optimized, and the anode temperature $T_A$ is fixed at $800\mathrm{K}$.}
\end{figure*}

For the bidirectional discharge (BD) model, where the space charge effect is considered in the presence of the bidirectional discharge, the output characteristics of PETE devices can be analyzed at different operating voltages. As shown in FIG.~\ref{fig:output}(a), when the operating voltage is relatively low, the net current density $J$ of the bidirectional discharge (BD) model (red lines) does not vary significantly with the operating voltage. This is because at low voltages, all electrons emitted from the cathode are within the forward accelerating electric field and can successfully traverse the electrode gap to reach the anode. As a result, the current density $J_C$ from the cathode [red lines in FIG.~\ref{fig:output}(c)] is much higher than the current density $J_A$ from the anode [red lines in FIG.~\ref{fig:output}(d)], and the overall current density is insensitive to the operating voltage. As the voltage increases, the electron concentration $n$ [FIG.~\ref{fig:output}(e)] in the conduction band of the cathode increases, and simultaneously, the maximum motive $\psi \left( {{x_m}} \right)$ [FIG.~\ref{fig:output}(f)] increases. Electrons from the cathode need to overcome the obstacle of the space-charge potential barrier in order to reach the anode. As a result, the net current density $J$ starts to decrease as the current density $J_C$ from the cathode decreases. For the anode, the potential barrier for electrons to surmount is lowered. This results in an increase in the anode current $J_A$. 

In FIG.~\ref{fig:output}, the blue lines correspond to the forward discharge (FD) model, where the space charge effect is analyzed without considering the reverse current $J_A$ from the anode. By comparing the BD model and the FD model, it is found that the higher the anode temperature $T_A$ is, the greater the difference between the two models. This is because as the anode temperature $T_A$ rises, the anode current $J_A$ increases as well, which in turn affects the space charge distribution and the potential distribution [FIG.~\ref{fig:output}(e) and (f)]. For example, in the case of the BD model, the higher the anode temperature $T_A$ is, the lower the cathode current density $J_C$. In contrast, for the FD model, the higher the anode temperature $T_A$ is, the higher the cathode current density $J_C$. These phenomena should be understood in the context of the electron recycling effect and the space charge effect. The electron concentration on the cathode surface increases with rising anode temperature for both the BD model and the FD model, but the space charge distribution and the potential distribution for the BD model exhibits more pronounced changes.

As shown in FIG.~\ref{fig:output}(b), when the operating voltage is low, the efficiency $\eta$ increases as the voltage is raised. After the efficiency $\eta$ reaches its maximum value, it then decreases as the voltage continues to increase. Additionally, the efficiency $\eta$ also decreases as the anode temperature $T_A$ increases, and the difference in efficiency values between the BD model and the FD model becomes larger and larger with increasing anode temperature. When the anode temperature $T_A$ is relatively high, in order to effectively evaluate the performance of the PETE device, the BD model needs to be considered.

FIG.~\ref{fig:anode property}(a) shows the efficiency $\eta$ of the BD model depends on the anode temperature $T_A$ and work function $\phi_A$, while FIG.~\ref{fig:anode property}(b) illustrates the deviation $\Delta \eta$ in efficiency between the FD model and the BD model. As can be seen from FIG.~\ref{fig:anode property}(a), the higher the anode temperature $T_A$ is, the lower the efficiency $\eta$. For example, when the anode work function $\phi_A$ is fixed at $0.9\mathrm{eV}$, $\eta$ will decrease from $24\%$ to $22\%$ as the anode temperature rises from $600\mathrm{K}$ to $800\mathrm{K}$. This is because according to Richardson's law [Eq.~\eqref{eq:J_SC}], a higher anode temperature leads to a greater reverse current generated by the anode. The reverse current will directly reduce the net current and impact the efficiency. On the other hand, as shown in FIG.~\ref{fig:anode property}(b), a higher anode temperature results in a larger $\Delta \eta$, since the reverse current has a noticeable impact on the space charge barrier in the BD model.

Additionally, as shown in FIG.~\ref{fig:anode property}(a), a smaller anode work function $\phi_A$ results in a greater efficiency $\eta$. For example, by fixing the anode temperature $T_A$ at $600\mathrm{K}$, $\eta$ will increase from $24\%$ to $28\%$ as the anode work function decreases from $0.9\mathrm{eV}$ to $0.6\mathrm{eV}$. This is because reducing the anode work function $\phi_A$ increases the voltage $V$ [Eq.~\eqref{eq:J_SA}] of the PETE device, enhancing the power output and thereby increasing efficiency. However, as shown in FIG.~\ref{fig:anode property}(b), a smaller anode work function $\phi_A$ results in a larger $\Delta \eta$. It is for the reason that a smaller anode work function $\phi_A$ makes it easier for electrons in the anode to overcome the potential barrier and be emitted, resulting in a larger anode reverse current $J_A$. By comparison with the FD model, the BD model takes into account the space charge generated by reverse current, leading to lower efficiency.

FIG.~\ref{fig:operating conditions} presents (a) the efficiency $\eta$ of the BD model and (b) the difference $\Delta \eta$ in efficiency between the FD model and the BD model varying with the gap width $d$ and the solar concentration $C$, where the voltage $V$ has been optimized. It is shown that there is an optimal solar concentration $C$ that maximizes the the efficiency for a given value of $d$. Initially, as $C$ increases, the temperature of the cathode also rises, which enhances the current density $J_C$ from the cathode and contributes to improved efficiency $\eta$. However, as $C$ continues to increase, the space charge accumulation effect becomes more pronounced, and the rate of the increase of excited electrons fails to keep pace with that of the photons, resulting in a decrease in efficiency $\eta$. FIG.~\ref{fig:operating conditions}(a) also shows that a larger gap width $d$ results in a lower maximum efficiency $\eta$. This indicates that using a smaller gap width $d$ can not only reduce the space charge effect but also allow the PETE device to better handle higher solar concentrations and achieve greater efficiency. 

As shown in FIG.~\ref{fig:operating conditions}(b), there is also an extreme value relationship between $\Delta \eta$ and the solar concentration ratio $C$. This is because when $C$ is small, the net current density $J$ is low, and the effect of the space charge is not significant. In addition, when $C$  is sufficiently large, the space charge potential barrier is mainly contributed by the cathode current density $J_C$, resulting in a small  $\Delta \eta$. It can also be seen in FIG.~\ref{fig:operating conditions}(b) that, with the increase of $d$, $\Delta \eta$ initially increases and then decreases. When $d$ is small, the device does not show significant space charge accumulation effects. On the other hand, when $d$ is large, the net current density decreases, leading to a reduction in the space charge accumulation effect, and the overall efficiency of the device is also lower. 

\section{\label{sec:Conclusions}Conclusions}
The current work proposes a novel approach to investigate the necessity and appropriate timing for incorporating the space charge potential barrier induced by bidirectional discharge into the theoretical computational simulations of PETE devices. We integrated the electron recycling effect, energy balance constraints, and space charge phenomena with the bidirectional discharge mechanism to obtain accurate predictions of the electron concentrations and cathode temperatures. The study found that the space charge effect induced by the reverse current can be significant when the anode temperature is high, the anode work function is low, and the solar concentration ratio and the gap width are at moderate levels. This approach provides a nuanced understanding of the significance of the space charge potential barrier arising from the bidirectional discharge phenomenon, and guides the refinement of computational models to achieve greater accuracy in evaluating the performance of PETE devices.

\begin{acknowledgments}
This work has been supported by the National Natural Science Foundation of China (12075197, 12364008, and 12365006), Natural Science Foundation of Fujian Province (2023J01006), Educational Teaching Reform Research Project of University Physics Discipline Alliance of Fujian Province (FJPHYS-2023-A02), Fundamental Research Fund for the Central Universities (20720240145), and Natural Science Foundation of Guangxi Province (2022GXNSFBA035636).
\end{acknowledgments}


\bibliography{PETE_SCBD_apssamp}

\end{document}